\documentstyle[12pt,epsfig]{article}
\begin{document}
\title{\vspace{-15mm}
       {\normalsize \hfill
       \begin{tabbing}
       \`\begin{tabular}{l}
	 SLAC--PUB--6471 \\
	 RI--2--94  \\
	 March 1994 \\
	 hep--th/9403179 \\
	 T   \\
	\end{tabular}
       \end{tabbing} }
       \vspace{8mm}
\setcounter{footnote}{1}
String--Kaluza--Klein Cosmology\thanks{Work
supported by the Department of Energy, contract DE--AC03--76SF00515.}}

\renewcommand{\thefootnote}{\fnsymbol{footnote}}
\vspace{5mm}

\author{
\setcounter{footnote}{2}
Klaus Behrndt\thanks{e-mail: behrndt@jupiter.slac.stanford.edu, Work
 supported by a grant of the DAAD} \\ {\normalsize \em
 Stanford Linear Accelerator Center} \\
{\normalsize \em Stanford University, Stanford, California 94309}\\[3mm]
\setcounter{footnote}{6} Stefan F\"orste
\thanks{e-mail: sforste@vms.huji.ac.il, Work supported by a grant
of MINERVA}\\ {\normalsize \em Racah Institute of Physics} \\
{\normalsize \em The Hebrew University, 91904 Jerusalem, Israel}}
\date{}

\maketitle

\begin{abstract}
We generalize a five dimensional black hole solution of low energy
effective string theory to arbitrary constant spatial curvature. After
interchanging the signature of time and radius we reduce the 5d
solution to four dimensions and obtain that way a four dimensional
isotropic cosmological space time.  The solution contains a dilaton,
modulus field and torsion. Several features of the solution are
discussed.
\end{abstract}

\renewcommand{\arraystretch}{2.0}
\renewcommand{\thefootnote}{\alph{footnote}}
\newcommand{\be}{\begin{equation}}
\newcommand{\ee}{\end{equation}}
\newcommand{\ba}{\begin{array}}
\newcommand{\ea}{\end{array}}
\newcommand{\vsf}{\vspace{5mm}}
\newcommand{\NP}[3]{{\em Nucl. Phys.}{ \bf B#1#2#3}}
\newcommand{\marpar}{\marginpar[!!!]{!!!}}

\section{Introduction}

One of the important motivations to discuss string theory is that
string theory contains quantum gravity. However, it is very
complicated to get results based on quantum gravitational effects
(like e.g.\ scattering of gravitons). Therefore a common and useful
way to obtain some insight in the fate of space time predicted by
string theory is to study the low energy effective string theory as a
classical theory and to look whether remarkable differences to
Einstein gravity occur. For example it seems to be more promising to
solve the problem of information loss in the process of collapsing
matter \cite{hawking} in low energy effective string theory (CGHS
gravity) rather than in Einstein gravity. Here the main reason for
the simplification due to string theory is that string theory enables
one to study that phenomenon in lower dimensions. (String theory
motivates 2d dilaton gravity whereas the Einstein--Hilbert action is
trivial in two dimensions.) A good review about
those problems is given in \cite{jef}.

Another subject where one might expect interesting stringy
modifications of general relativity is cosmology. Unfortunately, in
string cosmology there are not so many interesting solutions
available. One method to obtain exact solutions is given by
dealing with WZW or gauged WZW models. For gauged WZW models the
solution is actually not exact but the first order of an existing
exact solution.  Isotropic four dimensional cosmological solutions can
be obtained from an $SU(2)\times$Feigin--Fuchs WZW model \cite{ellis}.
However, because of the direct product structure one arrives always at
static cosmology. There are also exact non isotropic solutions
describing a non isotropic matter distribution \cite{amit}.

Since we are interested in dynamical isotropic cosmologies possibly
describing new scenarios for the universe we have to stick to the less
elegant phenomenological approach. That is, we will just consider
solutions of the low energy effective theory and lose that way
information about whether those solutions belong actually to a CFT.
However, there might be exact isotropic solutions with time dependent
world radius. Probably one can get those solutions by gauging WZW
models containing the $SU(2)$ as a subgroup.  The gauging should be
done in such a way that it does not affect the $SU(2)$ subgroup and
thus the $S_3$ structure survives. Since that is very involved we
choose the phenomenological approach for the time being and hope that
our solution corresponds to an (up to now unknown) CFT.  Later on we
will mention some facts which suggest that there is a CFT
corresponding to our five dimensional solution.

In a recent paper \cite{wir} we have obtained a four dimensional
cosmological solution containing a dilaton and a graviton.
There we started with a five dimensional black hole solution of
Einstein gravity and reduced it to a four dimensional cosmological
solution. In order to obtain an effective string action in four
dimensions we had to combine the reduction with a Weyl rescaling of
the metric. Thus the way through the five dimensional theory had a
purely technical meaning.  In the present paper we will take the five
dimensional origin of our solution more seriously, i.e.\ we will
combine string and Kaluza--Klein theory. That way we will be able to
incorporate also an antisymmetric tensor field.

If a higher dimensional solution of a theory does not depend on a
certain number of coordinates one can compactify those coordinates in
a Kaluza--Klein way \cite{schwarz}.  In our case we will take a
twisted static 5d black hole (BH) solution and will compactify the
coordinate which was the time before the twisting. (Twisting means
here that we interchange the signature of radius and time.)  The
resulting four dimensional solution will be regarded   as a
cosmological solution.  The degree of freedom corresponding to the 5th
coordinate leads to an additional scalar field in the 4d theory.

Before discussing the explicit solution let us briefly explain the
procedure in general.  Our starting point is the 5d string effective
action
\be   \label{1}
S^{(5)} = \int d^5 x \sqrt{G} e^{-2 \psi} \left( R + 4 (\partial \psi)^2
  - \frac{1}{12} H^2 \right)
\ee
where $\psi$ is the dilaton field and $H_{\mu\nu\lambda} =
\partial_{[\mu} B_{\nu\lambda]}$ is the torsion corresponding to the
antisymmetric tensor field $B_{\mu\nu}$. Since we are interested in a
4d cosmological interpretation we want to end up with a 4d
Friedmann--Robertson--Walker metric
\be   \label{2}
\tilde{ds}^2 = - d\tau^2 + a(\tau)^2 d\Omega_{k}^2
\ee
with $a(\tau)$ as world radius and $d\Omega_{k}^2$ as 3d volume
measure with the constant curvature $k$ (1,0,-1)\footnote{A possible
parameterization is: $d\Omega_{k}^2 = d\chi^2 + \left(\frac{\sin
\sqrt{k} \chi}{\sqrt{k}} \right)^2\left( d\theta^2 +
\sin^2\theta d\varphi^2 \right)$ .}.
The most general 5d metric respecting this 4d geometry is given (due to
Birkhoff's theorem) by a Schwarzschild solution
\be  \label{3}
ds^2 = \rho^2 dy^2 - e^{\lambda} dt^2 + t^2 d\Omega_{k}^2 =
 \rho^2 dy^2 + \tilde{G}_{\mu\nu} dx^{\mu} dx^{\nu}
\ee
where $y$ is the 5th coordinate. Assuming that $\rho$ and $\lambda$
are functions of $t$ only and furthermore that $\psi$ and $H$ are also
independent of $y$ (and $B_{\mu 5} = 0$) we can reduce (\ref{1}) to
\cite{schwarz}
\be  \label{4}
S^{(5)} \rightarrow S^{(4)} = \int d^4x \sqrt{\tilde{G}} e^{-2 \phi}
 \left( \tilde{R} + 4 (\partial\phi)^2 -
(\frac{\partial \rho}{\rho})^2 - \frac{1}{12} H^2 \right) \ .
\ee
After that the 4d metric is given by the ``spatial'' part of the
5d metric and the dilaton is
\begin{equation} \label{dil}
\phi = \frac{1}{4} \lambda = \psi - \frac{1}{2} \log \rho .
\end{equation}

This procedure to construct new cosmological solution has two
advantages. First, one can start with known static solutions (in
general BH solutions), generalize them to arbitrary spatial curvature
$k$ and after twisting and dimensional reduction one ends up with a
cosmological solution in four dimensions. That way it is easy to find
cosmological solutions even for non flat spatial curvature. Those
cosmological solutions contain matter described by the dilaton,
modulus and antisymmetric tensor field.  Secondly, because the 5d
theory has an abelian isometry (corresponding to the independence of
the 5th coordinate) it is possible to use the duality (or also O(d,d))
symmetry to construct new solutions. Note, that in standard (4d)
cosmology dualizing the isometries destroys the spherical symmetry.
Thus, especially for solutions with non flat spatial curvature the
Kaluza--Klein approach has some advantages.

In the literature several solutions have been discussed so far. But
only few complete analytic solutions are known.  For $k=1$ there is an
exact solution corresponding to the SU(2) WZW model \cite{ellis}. As
already mentioned this solution is static in the string frame and
becomes time dependent in the Einstein frame (via a Weyl
transformation containing the dilaton). For $k=0$ an analytic solution
was found by A.A.\ Tseytlin \cite{tseyt}. This solution has a time
dependent world radius in the string frame as well.  Examples for
anisotropic solutions which expand in different directions with
different velocities are given in \cite{tseyt3,luest} (see also
references therin).  In this case a couple of solutions correspond to
gauged WZW theories. Cosmological scenarios coming from string theory
are, e.g., discussed in \cite{tseyt3} (with special emphasis on the
role played by the dilaton) and in \cite{venz} (inflation and
deflation scenarios). Numerical results showed that the initial
conditions have a significant influence on different scenarios
\cite{perry}. Our 4d theory differs from these solutions by an
additional scalar field (modulus) and a vanishing central charge term
but via the Kaluza--Klein approach we were able to find an analytic
solution for arbitrary $k$. In the context of Einstein gravity
(vanishing dilaton) the Kaluza--Klein approach via 5d BH has been
discussed in \cite{matzner,wltsh,gibbons}. Here, via the inclusion of
a cosmological constant it was possible to get a period of exponential
inflation \cite{wltsh}. But to obtain a sufficient long inflation an
extremal accuracy of the initial conditions is necessary.

We have organized the paper as follows.  In the next section we want
to start with an explicit 5d solution and consider the properties of
this solution. In the third section we will discuss the effective 4d
theory and the cosmological interpretation of this solution. A forth
section is devoted to the asymptotic behavior, the behavior of the
dilaton and modulus field, and to the discussion of singularities.
Finally we will give some conclusions.

%----------------------------------------------------------------------
\section{Five dimensional solution}
\noindent From (\ref{3}) follows that we have to look for a black 
hole solution of the 5d theory (\ref{1}). Fortunately, this problem
was already solved by Gibbons and Maeda \cite{maeda}. Based on this
solution Horowitz and Strominger developed a black p-brane solution
\cite{strom}.  It is not very difficult to generalize their solution
to arbitrary constant spatial curvature $k$ and one gets
\be  \label{5}
\ba{c}
ds^2 = \frac{-k+\left(\frac{t_+}{t}\right)^2}{1-\left(\frac{t_-}{t}\right)^2}
     dy^2 - \frac{1}{(-k+(\frac{t_+}{t})^2)(1-(\frac{t_-}{t})^2)} dt^2
    + t^2 d\Omega_{k}^2 \\
e^{-2(\psi - \psi_{0})} = 1 - \left(\frac{t_-}{t}\right)^2 \qquad , \qquad
  H = 2 t_+ t_- \epsilon_{3,k}
\ea
\ee
where in the torsion $2 t_- t_+ =Q_M$ defines a magnetic charge and
$\epsilon_{3,k}$ is the volume form corresponding to
$d\Omega_k$, i.e.\ $$ \epsilon_{3,k}=\left( \frac{\sin \sqrt{k}
\chi}{\sqrt{k}}\right)^2
\sin \theta \, d\chi \wedge d\theta \wedge d\varphi $$

Note, that in this (cosmological) context the time $t$ is the variable
playing the role of the radius of the 3-sphere (e.g. for $k=+1$) and
that in comparison to the usual BH physics the signature is different.
If we set $k=1$, interpret $y$ as time and $t$ as radius this is just
the solution in the notation of \cite{strom}. It has a curvature
singularity at $t^2 =t_{-}^2$ and at $t=0$ and an event horizon at
$t^2=t_+^2$.  It is interesting that this solution corresponds
%to the first order of an exact
to a conformal field theory in the limit $t_- =t_+
$ for $k=1$ \cite{gidd}. In the following we will give an argument
that also for arbitrary $k$ there is such a limit.  It is useful to
transform the solution to the conformal time $\eta$ (from the point of
view of 4d cosmology).  Assuming that $t_{+}^2 > k t_{-}^2$ this
transformation is given by
\be \label{501}
t^2 = t_{-}^2 + ( t_{+}^2 - k t_{-}^2) \left(\frac{\sin \sqrt{k} \eta}
 {\sqrt{k}}\right)^2 \ .
\ee
For the solution (\ref{5}) we obtain
\be \label{502}
\ba{l}
ds^2 = \left( \frac{\sqrt{k}}{\tan \sqrt{k} \eta } \right)^2 \, dy^2
  + \left\{ t_{-}^2 + (t_{+}^2 - k t_{-}^2) \left(\frac{\sin \sqrt{k}
  \eta }{\sqrt{k}}\right)^2 \right\} \, \left[ -d\eta^2 +
  d\Omega_{k}^2 \right] \\
 e^{2 (\psi - \psi_0)} = 1 + \frac{t_{-}^2}{(t_{+}^2 - k t_{-}^2)
 \left(\frac{\sin \sqrt{k} \eta }{\sqrt{k}}\right)^2}
\ea
\ee
and the torsion remains unchanged. In this parameterization it can be
seen that for $k=1$ the ``spatial'' part oscillates between the minimal
extension $t_-^2$ and the maximal extension $t_+^2$ and at these extrema
the compactification radius of the 5th coordinate $\rho$ has a pole or
zero, respectively. Similar to \cite{gidd} we are interested in a
critical limit in which the spherical coordinates decouple from the
$(y,\eta)$ part. An obvious possibility is that $t_{+}^2 = k t_{-}^2$.
In this limit our 5d theory decouples in a direct product of a 3d
(spherical) part and the 2d ($\eta , y$) part.  The (constant)
spherical part of the metric is then given by $t_{-}^2 d\Omega_{k}^2$
and the torsion by (\ref{5}).  In order to prove the conformal
invariance of this 3d theory we calculate the generalized curvature
$\hat{R}$ which is defined in terms of the connection $\hat{\Gamma} =
\Gamma + \frac{1}{2} H$ (where $\Gamma$ is the standard Christoffel
connection)
\be \label{503}
\hat{R}_{mnab}=R_{mnab}+
   \frac{1}{2} \left( D_{b} H_{mna} - D_{a}
   H_{mnb}\right) + \frac{1}{4}\left(H_{mbr}
   H^{r}_{~na} - H_{mar}  H^{r}_{~nb}\right)
\ee
($m,n,a,..$ are the indices of the 3d theory) and we get the result
\be  \label{504}
\hat{R}_{mnab} = \left( \frac{k}{t_{-}^2} - \frac{t_{+}^2}
 {t_{-}^4} \right) \, \left( G_{ma} G_{nb} \, - \, G_{mb}
G_{na} \right) \ .
\ee
Thus, in the limit $t_{+}^2 = k t_{-}^2$ this curvature vanishes which
means that the 3d subspace is parallelized and therefore the
corresponding $\sigma$ model  is conformally invariant \cite{braaten}.
But because the dilaton is divergent in this limit we have to shift
the constant part $\psi_0$ simultaneously. The complete extremal limit is
therefore given by
\be \label{505}
\psi_0 \rightarrow \psi_0 + \frac{1}{2} \log (t_{+}^2 - k t_{-}^2)
\qquad \mbox{and then} \qquad  t_{+}^2 \rightarrow k t_{-}^2
\ee
and the resulting 5d metric and dilaton are
\be  \label{506}
\ba{l}
ds^2 = \left( \frac{\sqrt{k}}{\tan \sqrt{k} \eta } \right)^2 \, dy^2
  + t_{-}^2  \left[ -d\eta^2 + d\Omega_{k}^2 \right] \\
 e^{2 (\psi - \psi_0)} = \frac{t_{-}^2}
{\left(\frac{\sin \sqrt{k} \eta }{\sqrt{k}}\right)^2} \qquad , \qquad
 H = 2 \sqrt{\pm k} t_{-}^2 \epsilon_{3,k} \ .
\ea
\ee
Now, we are going to discuss the three cases ($k=-1,0,1$) separately.
For $k=1$ the spherical part is a $S_3$ manifold and the conformal
field theory resulting from the vanishing of $\hat{R}$ (\ref{503}) is
the SU(2) WZW theory which is only consistent if the radius of the
$S_3$ is quantized \cite{witt2,ellis} $t_{-}^2 = 1,2,3,..$.  After a
Wick rotation in $\eta$ the remaining 2d theory ($(y,\eta)$ part of
the metric and the dilaton) is the dualized 2d black hole solution
corresponding to an $SL(2,R)/U(1)$ coset model \cite{witt,amit2,verl}.
If we scale $y$ ($y \rightarrow y t_{-} $) we find that the levels of
both WZW models are given by $t_{-}$ and consequently the level of the
gauged WZW model leading to the 2d black hole is quantized, too.

For $k=-1$ (remember, here $k$ is  {\bf not} the level of the WZW model
but the spatial curvature) the 3-space is a pseudo sphere and the 2d
part is the standard $SL(2,R)/SO(1,1)$ 2d black hole. Again, $t_{-}^2$
corresponds to the level of the 2d black hole coset model, but now,
because the 3d part corresponds not to a compact group\footnote{We do
not know whether behind this theory is a group at all.} we have no
quantization condition for $t_{-}$. For vanishing $k$  the situation
is a little  curious.  Below, we will give the duality transformation
(\ref{7}) leading to a new solution. If we perform that transformation
in (\ref{506}) and set $k=0$ we find that the 2d part is just the 2d
Minkowski space written in polar coordinates and since the spherical
part is flat as well we have a 5d Minkowski space. Furthermore the 5d
dual dilaton for $k=0$  vanishes in this limit, too. Therefore, for
$k=0$ the extremal limit yields a (dualized) trivial 5d theory.

Before we turn back to our original parameterization (\ref{5}) we want
to discuss another extremal limit. The above extremal limit
(\ref{505}) for $k=1$ makes sense only in the conformal time. In our
original time (\ref{5}) we would get a wrong sign of $G_{tt}$
or equivalently the allowed $t$-region would shrink to zero (see
below).  But for $k=-1$ we can perform the extremal limit first and
afterwards transform it to the conformal time. The result is
\be  \label{507}
ds^2 = dy^2 + \left\{\frac{1}{2} e^{2 \eta} + t_{-}^2 \right\}
  \left[ -d\eta^2  +  d\Omega_{k=-1}^2 \right] \ .
\ee
If we perform a further time rescalling in order to get a conformal
flat spatial part we find
\be \label{508}
\ba{l}
ds^2 = dy^2 + \left\{1 + \frac{t_{-}^2}{\eta^2} \right\}
  \left[ -d\eta^2 + \eta^2 d\Omega_{k=-1}^2 \right] \\ e^{2 \psi} = 1
+ \frac{t_{-}^2}{\eta^2} \ .
\ea \ee
In this conformal limit the compactification radius of the 5th
coordinate is constant \mbox{($\rho=1$)} and the remaining part is
conformally equivalent to the 4d Minkowski space with the conformal
factor given by the dilaton field. The geometry of this limit is $R\,
\times \,$4d throat, see figure (2d). The conformal exactness of this
model is not due to a correspondence to a WZW model.  Instead, it is
possible to show that this model has a super symmetric generalization
((4,4) extended world sheet) for which a non-renormalization theorem
exists \cite{call}. The spatial parts of both extremal 5d theories
(\ref{506}) and (\ref{508}) are different.  The former one is constant
in time and has no flat regions whereas the later one has a time
evolution and one flat limit (for $\eta \rightarrow \infty$). We will
come to this point again when we discuss the effective 4d theory.

Let us now turn back to our original time parameterization (\ref{5}).
Since none of the background fields in (\ref{5}) depends on $y$ we can
apply a duality transformation with respect to  the $y$ direction
\cite{busch},
\begin{equation}   \label{7}
\begin{array}{c c l}
G_{55} & \longrightarrow  &\frac{1}{G_{55}} \\
\psi & \longrightarrow & \psi -\frac{1}{2}\log G_{55}
\end{array}.\end{equation}
Therewith the dual solution is given by
\begin{equation}
ds^2 = \frac{1-\left( \frac{t_-}{t}\right)^2}{-k+\left(\frac{t_+}
      {t}\right)^2}
dy^2 - \frac{dt^2}{\left(-k+\left(\frac{t_+}{t}\right)^2\right)
\left(1 -\left(\frac{t_-}{t}\right)^2\right)}
+t^2 d\Omega_k^2
\end{equation}
\begin{equation}
e^{-2\psi} \sim -k+\left( \frac{t_+}{t}\right)^2\;\;\;\; , \; \; \; \;
H=2 \, t_+t_- \epsilon_{3,k}.
\end{equation}
From(\ref{3}) and (\ref{dil}) it is clear that in the 4d theory (after the
compactification has been performed) the duality transformation
(\ref{7}) results only in an inversion of the compactification radius
(modulus field),
\begin{equation} \label{dtrafo}
\rho \longrightarrow \frac{1}{\rho},
\end{equation}
which is a manifest symmetry in (\ref{4}). The 4d dilaton is not
affected by the 5d duality transformation.

Before we discuss the effective 4d theory it is worthwhile to consider
the special case of one vanishing constant ($t_{+}$ or $t_{-}$).  (A
vanishing of both makes sense only for $k= -1$ yielding a 5d flat
theory (cf. (\ref{5})).) In either case the torsion vanishes.  If we
set $t_{-}=0$ the result is the standard 5d black hole generalized to
arbitrary constant spatial curvature.  The other case ($t_{+}=0$)
defines a 5d dilaton graviton system, but the resulting 5d black hole
solution is just the dualized standard 5d black hole \cite{wir2}.  In
that sense both cases are dual to each other, and thus, it is
sufficient to consider only one case, e.g.\ $t_{-}=0$. In this case we
have a 5d Einstein-Hilbert theory without a dilaton field. Therefore,
if we reduce this 5d theory only one scalar field coming from $G_{55}$
arises in the effective 4d theory.  That leads to a further option of
interpretation. Namely we can perform an additional Weyl
transformation in such a way that we obtain a 4d string effective
action with a dilaton and a graviton only \cite{wir}.
%----------------------------------------------------------------------------
\section{Effective 4 dimensional theory}
After we have given a solution of the 5d theory we are ready to study the
Kaluza--Klein reduced 4d theory. A stationary point of (\ref{4}) is given
by
\begin{equation} \label{4d}
\begin{array}{c c l}
ds^2 & = &  -\frac{dt^2}{\left( -k+\left(\frac{t_+}{t}\right)^2\right)
\left( 1-\left(\frac{t_-}{t}\right)^2\right)}+t^2d\Omega_k^2 \\
\rho^2 & = &\frac{-k+\left(\frac{t_+}{t}\right)^2}{1-\left(\frac{t_-}{t}
\right)^2} \\
e^{-2\phi} & \sim & \sqrt{ \left( 1-\left(\frac{t_-}{t}\right)^2\right)
\left( -k+\left(\frac{t_+}{t}\right)^2\right)}
\end{array}. \end{equation}
In order to have a well defined metric in (\ref{4d}) we have to restrict
the time $t$ to those regions where the $G_{tt}$ component is greater
than zero,
\begin{equation}   \label{int}
\left( -k +\left(\frac{t_+}{t}\right)^2\right)
\left(1 -\left(\frac{t_-}{t}\right)^2\right) > 0.
\end{equation}
Hence both factors in (\ref{int}) have to be either positive or negative.
This leads to the two possibilities
$$\begin{array}{l c r}
t_+^2 > k t^2 \;\;&\;\; t_-^2 < t^2 & \qquad \qquad (a)\\
\mbox{or} & & \\
t_+^2 < kt^2 \;\;&\;\; t_-^2 > t^2 & \qquad \qquad (b)
\end{array}. $$
Since these restrictions are very important for the fate of the
universe we will discuss them in a more detailed way. The time $t$ in
(\ref{4d}) defines the radius of the 3 space and therefore these
conditions restrict the region for the world radius (and not the life
time as one might think).  A bounded $t$ region results in a bounded
three space. In order to get a better impression we transform the
solution to the conformal time coordinate $\eta$ defined by
(\ref{501}). For (\ref{4d}) we find
\be \label{4dconf}
ds^2 = \left\{ t_{-}^2 + (t_{+}^2 - k t_{-}^2) \left(\frac{\sin \sqrt{k}
  \eta }{\sqrt{k}}\right)^2 \right\} \, \left[ -d\eta^2 +
  d\Omega_{k}^2 \right] \ .
\ee
First, we consider the compact case $k=1$ where $t$ is simply the
radius of the three sphere. In that case from (a), (b) follows
$$ \begin{array}{ c c l r}
t_-^2 < & t^2 &<t_+^2,& \qquad \qquad (a^{\prime})\\
t_+^2 < & t^2 & <t_-^2, & \qquad \qquad (b^{\prime})
\end{array}$$
respectively. That means that the three space oscillates between the
extrema $t_{\pm}$. However, that smooth interpretation of the 4d
metric is only possible for $t_{-}^2>0$ (let us assume that $t_-^2 <
t_+^2$). Otherwise for $t_-^2 <0$ the radius of the 3 sphere shrinks
to zero at $t=0$ and the corresponding singularity is the beginning of
the universe.  Moreover, the universe ends with a singularity where
the Weyl factor in (\ref{4dconf}) vanishes again. In this case the
lifetime of the universe is finite whereas in the former case we have
an eternal oscillation (see figure (1a,2a)).  For $k=0, -1$ the
solution is not oscillating and if $t^2_{\pm}>0$ the $t$ region is
infinite but bounded from below. Again, if one constant is negative ,
e.g. $t_-^2 <0$, from (a) and (b) follows that the $t$ region could be
finite.

So far we have considered our solution in the string frame, i.e.\ in
the frame wherein extended objects propagate. However, another
description is possible in the Einstein frame which is given by a
redefinition of the metric
\begin{equation}
G_{\mu \nu}^{(E)} = e^{-2\phi}\, G_{\mu \nu} \ .
\end{equation}
After this redefinition the effective action (\ref{4}) becomes
\begin{equation}
S=\int d^4x\sqrt{G^{(E)}}\left[ R^{(E)}-2(\partial \phi)^2 -
\left(\frac{\partial\rho}{\rho}\right)^2 -\frac{1}{12}
e^{-4 \phi}H^2\right]
\end{equation}
that contains the standard Einstein--Hilbert part. In the Einstein
frame our 4d solution takes the form
\begin{equation}
ds_E^2 = -\frac{dt^2}{\sqrt{\left( -k + \frac{t_+^2}{t^2}\right)
\left( 1 -\frac{t_-^2}{t^2}\right)} } +
t^2 \sqrt{\left( -k +\frac{t_+^2}{t^2} \right) \left( 1-\frac{t_-^2}{t^2}
\right)} d\Omega_k^2.
\end{equation}
(The other fields are the same in the Einstein and String frame.) In
the following we will consider both frames on an equal footing.  There
are several arguments in favour of one or the other frame.  From the
string $\sigma$ model follows, that the string couples to the string
metric and that the free motion of a string follows geodesics in the
string frame \cite{bell}. On the other hand if one wants to get the
string correlation functions with correct vertex operators from the
string $\sigma$ model one has to take the Einstein frame (e.g. see
\cite{tseyt2}). But all these arguments are not compelling for one
or the other frame. Instead, if one measures, e.g. the world radius,
in some physical units defined by the Compton wave length of matter
contributions both frames yield the same results \cite{linde}.

Up to now we have used the coordinate system from the 5d black
hole solution. But that is not the convenient system for
the discussion of cosmological scenarios. Usually one takes the
Friedmann--Robertson--Walker system defined by the metric
\begin{equation}    \label{robby}
ds^2 = -d\tau ^2 + a^2 (\tau) d\Omega_k^2 .
\end{equation}
The corresponding coordinate transformation is given by the solution
$t(\tau)$ of the following differential equations
\begin{equation}  \label{strdgln}
\dot{t}^2(\tau) = \left(-k +\frac{t_+^2}{t^2}\right)
\left(1-\frac{t_-^2}{t^2}\right)
\end{equation}
in the string frame and
\begin{equation} \label{edgln}
\dot{t}^2(\tau_E) = \sqrt{\left(-k+\frac{t_+^2}{t^2}
\right) \left( 1-\frac{t_-^2}{t^2}\right)}
\end{equation}
in the Einstein frame. In the string frame the world radius $a$ is given by
\begin{equation} \label{wrad}
a^2 (\tau ) = t^2 (\tau )
\end{equation}
whereas in the Einstein frame it gets an additional factor
\begin{equation} \label{wrade}
a_E^2(\tau_E ) = t^2(\tau_E )\sqrt{\left( -k+\frac{t_+^2}
{t^2(\tau_E )}\right)
\left( 1- \frac{t_-^2}{t^2(\tau_E )}\right) }.
  \end{equation}
  \begin{figure}[t]
%\vspace*{-30mm}
\vspace*{-40mm}
\begin{tabular}{c}
  \par \hfill \begin{minipage}[t]{155mm}
  \mbox{\epsfig{file=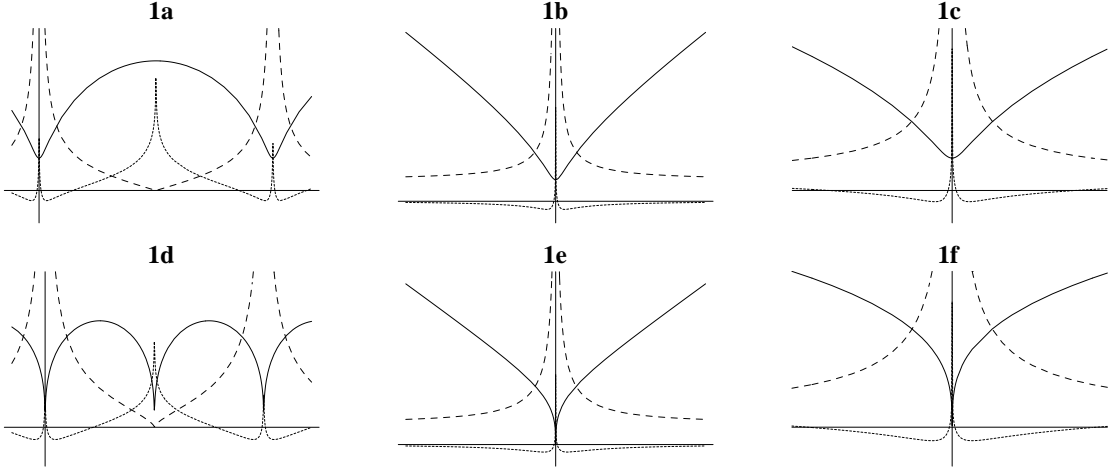,width=160mm}} \hfill
%\vspace{-20mm}
\vspace{-50mm}
  \par
    \caption{Plots for the world radius $a(\tau )$, the dilaton $\phi(\tau)$
 (doted) and the modulus $\rho(\tau)$ (dashed) in the string frame (a,b,c)
  and in the Einstein frame (d,e,f). On the left hand side is $k=1$, in the
  middle $k=-1$ and on the right hand side is $k=0$. All Plots are
  calculated for $t_+ =4$ and $t_- =1$}
     \end{minipage} \hfill
%\vspace*{0mm}
\vspace*{-15mm}
\end{tabular}
\end{figure}
Unfortunately we are not able to find an analytic solution for
(\ref{strdgln}) and (\ref{edgln}). Only in some special cases for
vanishing constants or in the extremal limit we find an analytic
expression. In general we can perform the coordinate transformations
(\ref{strdgln}) and (\ref{edgln}) numerically only. In the following
we will discuss some plotted results.  Let us first assume that
$0<t_{-}^2 < t_+^2$. In figure (1a) we have plotted the world radius
for $k=1$. The doted line shows the dilaton behavior and the dashed
line the modulus field $\rho$. In this case the universe oscillates
between the minimum $t_-$ and the maximum $t_+$. At these points the
dilaton field is divergent and the modulus field is infinite at the
minima and zero at the maxima. Note, that the duality transformation
(\ref{dtrafo}) inverts that behavior, i.e. the modulus field is zero
at the minima and infinite at the maxima.  Figure (1d) shows the
same configuration but in the Einstein frame. Here, the universe
starts and ends with a singularity where the world radius vanishes and
the dilaton is divergent. The other figures correspond to $k=-1$
(1b,1e) and $k=0$ (1c,1f). The qualitative behavior is similar. In the
string frame (1c,1f) the universe is shrinking up to a minimum given
by $t_-$ (at $\tau=0$) and then expands for ever. In the Einstein
frame (1e,1f) we have again a singularity at this minimum and in both
frames the dilaton and modulus is divergent at $\tau=0$. The only
difference is given by the behavior at infinity. But this question we
are going to discuss in the next section.

For the present, we want to consider the case $t_- =0$ in the string
frame as an illustrative example.  There are two different
possibilities to handle this case. Firstly, we can take the solution
(\ref{4d}) set $t_-=0$ and then transform the solution to the proper
time $\tau$.  After that one gets a completely analytical expression
for $a(\tau)$.  The other possibility takes into account that $t_- =0$
corresponds to a vanishing 5d dilaton and torsion (cf.\ (\ref{5}))
resulting in a 5d Einstein--Hilbert theory.  Therefore, the 4d theory
can contain only one independent scalar field corresponding to
$G_{55}$ and the reduction procedure (\ref{4}), (\ref{dil}) can be
modified in order to end up with a 4d effective action (string or
Einstein) which contains only one scalar field. This procedure is
described in \cite{wir2} and the difference to the solution (\ref{4d})
with $t_-=0$ is given by an additional Weyl transformation.  But let
us describe the case $t_- =0$ for the present solution.  For $k=1$ we
will find the same results as Matzner and Mezzacappa \cite{matzner}.
The solution of (\ref{strdgln}) is given by\footnote{For $k=0$ it is
reasonable to restrict $\tau$ to positive values.}
\begin{equation} \begin{array}{ lc r } \label{matzn}
\tau = \pm k\sqrt{-kt^2 +t_+^2} & \qquad \mbox{for}\qquad & k=\pm 1\\
\tau =\frac{t^2}{2 |t_+|} & \qquad \mbox{for} \qquad & k=0 \ .
\end{array} \end{equation}
This leads to the following equations for the world radius $a$
(\ref{wrad})
\begin{equation} \begin{array}{lcr}
\tau^2 + k a^2 = t_+^2 & \qquad \mbox{for} \qquad & k=\pm 1\\
a=\sqrt{2|t_+|\tau} & \qquad \mbox{for} \qquad & k=0 \ .
\end{array} \end{equation}
For $k=1$ we get a half circle in the $(a,\tau )$ plane ($a>0$).  For
$k=-1$ the solution describes a hyperbolae in the $(a,\tau)$ plane.
Asymptotically we get the flat solution ($a=\tau$) with the restrictions
$a^2 < \tau^2$ for $t_+^2 >0$ or $a^2 > \tau^2$ for $t_+^2 < 0$.
%--------------------------------------------------------
\newpage
\section{Asymptotic behavior, Dilaton, Singularities ...}
In this section we discuss some special questions
in detail.
\vspace{3mm}

\noindent {\bf Asymptotic behavior:}
Although, it is not possible to solve the equations (\ref{strdgln})
and (\ref{edgln}) generally we can find analytic results in special
regions. First we want to consider the case $t(\tau) \rightarrow t_-$.
For $0 < t_- < t_+$ ($t_-$ and $t_+$ real) that limit corresponds to
minima in figure (1a-1c) and we find
\be  \label{s0}
\ba{ccc}
a(\tau ) \sim \tau ^2 + const. &\quad , \quad &
a_{E}(\tau ) \sim \tau ^{\frac{2}{3}}\\
e^{2\phi}\sim\tau^{-2}     &\quad , \quad &
e^{2\phi_{E}}\sim\tau^{-\frac{2}{3}} ~.
\ea
\ee
We see that in the string frame we have a quadratic time dependence of
the world radius.  In the Einstein frame $a(\tau)$ vanishes at $t(\tau
) = t_-$.  For $k=1$ we get the same behavior near the maxima in
figure (1a)  ($a\sim -\tau^2 + const$ or $a_E \sim (\tau_0 -
\tau)^{2/3}$).  This different behavior of the string frame in
comparison to the Einstein frame is caused by the divergence of the
dilaton near the extrema. A complete other behavior occurs if
$t_-^2<0$.  Then, $t=0$ is the lower bound and we obtain for $\tau
\rightarrow 0$ (or $t \rightarrow 0$)
\be  \label{e0}
\ba{ccc}
a(\tau ) \sim \tau ^{\frac{1}{3}} &\quad , \quad &
a_{E}(\tau ) \sim \tau ^{\frac{2}{3}}\\
e^{2\phi}\sim\tau^{\frac{2}{3}}     &\quad , \quad &
e^{2\phi_{E}}\sim\tau ~.
\ea
\ee
Again, for $k=1$ the world radius approaches zero with the same power
at the end of the universe. The qualitative feature in the string frame is
then the same as in the Einstein frame: the universe starts and ends with
a singularity ($a=0$).

The other asymptotic limit $\tau \rightarrow \infty$ or $t \rightarrow
\infty$ is possible for $k=0$ or $k=-1$ only (cf. (\ref{int})) and we
find
\be                                    \label{infty}
\ba{ccc}
a(\tau ) \sim \left\{
\ba{cc}
\sqrt{\tau} & \quad k=0 \\
\tau & \quad k=-1 \ea \right.
 & \quad , \quad &
a_{E} \sim \left\{
\ba{cc}
\tau^{\frac{1}{3}} & \quad k=0 \\
\tau & \quad k=-1 \ea \right. \\
e^{2\phi}\sim \left\{
\ba{cc} \tau^{\frac{1}{2}} & \quad k=0\\
1 & \quad k=-1 \ea \right.
   & \quad , \quad &
e^{2\phi_{E}}\sim \left\{
\ba{cc} \tau^{\frac{2}{3}}& \quad k=0 \\
1 & \quad k=-1 \ea \right.  ~.
\ea
\ee
This behavior does not depend on whether $t_-^2$ is greater or less
than zero. In the string frame there are no reasons to restrict $\tau$
to positive values. For negative $\tau$ we obtain the same behavior.
Remarkable, for $k=-1$ we have then in the infinite past and infinite
future two asymptotic flat regions and because $a(\tau)$
remains finite at $\tau=0$ (for $t_-^2>0$) these two flat regions are
connected by a wormhole (figure (2b)). For $k=0$ we have two
asymptotic non flat region which are connected. If we come from minus
infinity the universe shrinks down to a minimal size ($a=t_-$) and then
expands forever. But both regions are only connected in the string
frame. In the Einstein frame there is a curvature singularity
($a_E(\tau=0)=0$) between both regions.

Let us discuss the dilaton behavior. As it can bee seen in the figures
(1) and in the asymptotic behavior the dilaton is always divergent at
the extrema in the string frame. This can be explained as follows. If
we use (\ref{strdgln}) and (\ref{4d}) we get for $\dot{a}(\tau)$
($a(\tau) = t(\tau)$)
\be
\dot{a}\left(t(\tau)\right) = a'(t) \, \dot{t}(\tau) =  \sqrt{\left(-k +
\frac{t_+^2}{t^2}\right) \left(1-\frac{t_-^2}{t^2}\right)} = e^{-2\phi}
\ee
and thus $\phi=\infty$ at $\dot{a}=0$. Since the string and Einstein metric
differ by the conformal factor $e^{-2\phi}$ every extremum in the
string frame corresponds to a zero of $a_E$ in the Einstein frame
(cf.\ figure (1a) and (1d)).

We obtain similar results for the modulus field $\rho$ (cf.\
(\ref{4d})).  It is either zero, e.g. at $t^2 = t_+^2$ for $k=1$
corresponding to the maxima in figure (1a) or it is divergent at $t^2
= t_-^2$ corresponding to the minima. In contrast to the dilaton field
the modulus field is not uniquely determined. The reason is that we
can always invert the modulus by the duality transformation (\ref{7}).
The singularities of the scalar fields are somehow strange because at
these points the 4d string metric is smooth. Therefore we want to give
a detailed discussion on that singularities.

\vspace{2mm}

\noindent {\bf Singularities:}
In order to exclude pure coordinate singularities we calculate the
scalar curvature.  In the Robertson--Walker frame (\ref{robby}) the
curvature is given by \cite{wald}
\begin{equation}
R= 6\left(\frac{ \ddot{a}}{a} + \left( \frac{\dot{a}}{a}\right)^2
+\frac{k}{a^2} \right)
\end{equation}
where $\dot{a} = a'(t)\, \dot{t}$. If we use (\ref{strdgln})
and (\ref{wrad}) for the string frame  we find
\begin{equation}
R= 6 \, \frac{t_+^2 t_-^2}{t^6} \
\end{equation}
and for the Einstein frame we have to perform another transformation
(\ref{edgln}) in (\ref{wrade}) with the result
\begin{equation}   \label{re}
R = - \frac{3(t_+^2 - k t_-^2)^2}{2 [ (t^2 -t_-^2)
(t_+^2 - kt^2)]^{\frac{3}{2}}} \ .
\end{equation}
We observe that in the string frame there is only a singularity at $t =0$.
Whenever $t=0$  belongs to the allowed region the singularity has to
be considered as a big bang or big crunch singularity. If $t=0$ is not
in the allowed region there is no singularity at the edges of the $t$
region and hence an analytical continuation (e.g. by transforming to
the conformal time $\eta$) of the solution is  possible.
In the Einstein frame we have singularities at $t^2 =t_\pm^2 $
(as long as these points belong to the allowed $t$ region).

Let us discuss the situation for $k=1$. For $k=0,-1$ the arguments are
qualitatively the same. First let us assume that both constants are
positive, $t_{\pm}^2>0$ and that $t_+^2>t_-^2$. Whereas in the string
frame these values define only the maximal/minimal radius of the three
sphere and are non singular the situation is different in the Einstein
frame. In the Einstein frame, the radius of the three sphere contains
an additional square root which vanishes at $t^2=t_{\pm}^2$ and
therefore there are singularities at these points (cf.\ (\ref{wrade})). In
the other case where $t_-^2 < 0$ the string frame is singular at $t=0$
but in the Einstein frame the limit $t\rightarrow 0$ yields a finite
radius of the three sphere: $\sqrt{t_+^2(-t_-^2)}$ ($t_-^2<0$) which
is just the magnetic charge (see({\ref{5})) and therefore there is no
singularity at this point.  The same happens at $t^2=t_+^2$ but now
the Einstein frame is singular and the string frame is smooth. So, we
have the result that both frames are in some sense complementary with
respect to the (4d) space time singularities.  If one frame is
singular at one of the critical points ($t^2 = 0,t_-^2,t_+^2$) than in
the other frame there is no singularity at that point and vice verse.
The reason is that the dilaton is divergent at all these critical
points.  However, this conclusion is only possible as long as both
constants are non vanishing. If one constant vanishes than both frames
have simultaneously singularities, e.g. for $t_-=0$ (\ref{matzn}).

It is illustrative to study the singularity structure for the solvable
case $t_- =0$. For $k=1$ the point $t=0$  (= vanishing 3
sphere) is mapped on two points $\tau =\pm t_+ $. These are the two
end points of the half circle in the $(a,\tau )$ plane and hence there
is a big bang and a big crunch singularity. For $k=0$ we have a big
bang singularity at $\tau =0$.  For $k=-1$ and $t_+^2 < 0$ the point
$t=0$ is not in the allowed region (i.e.\ the universe shrinks not to
zero) and the allowed $\tau$ region is the real
axis. Hence there are no singularities whereas for $t_+^2 >0$ there
are singularities at $\tau = \pm t_+$. Cutting off the region where
$\tau $ or $a$ are less than zero we get a big bang singularity at
$\tau =|t_+|$.

Nevertheless, although the string metric for $t_{\pm}^2>0$ is non
singular the scalar fields are singular at the extrema of the string
frame. These singularities can be understood as remnants of the 5d
theory which has singularities there. But is the theory there really
singular? Let us explain the situation for $k=-1$.  Near the minimum,
i.e.\ for vanishing $\tau$ (or $\eta$) or inside the wormhole (see
figure (1b,2b)) the spherical part of the 5d metric decouples and the
2d part behaves like a dualized Lorentzian black hole (see (\ref{502})).
The spherical part is non singular there but the 2d part has a
singularity. On the other hand dualizing the theory yields the
standard 2d black hole which has no singularity at $\tau=0$ (only a
coordinate singularity corresponding to a horizon).  This is a
consequence of the known fact that for 2d stringy black holes the
target space duality transformation interchanges the horizon with the
singularity \cite{amit2,verl}. The question is now whether matter or
information can pass the wormhole. In \cite{verl} it was shown that
for winding modes it is possible to define vertex operators which are
regular even at the BH singularity. Thus, we can conclude that winding
modes can pass the wormhole, and furthermore, that here is an
essential difference to ordinary field theory: only string states can
pass the wormhole singularity.  In addition, the duality symmetry
which transforms winding modes to momentum modes ensures that both
modes are physically equivalent. This region in our cosmological
solution seems to be very interesting and deserves still further
investigations.

\vspace{2mm}

\noindent {\bf Extremal limit:}
In section two we pointed out that similar to the 5d black hole
solution it is possible to find a limit in which the 5d solution
corresponds to a conformal field theory. In that limit (\ref{505}) the
5d solution is given either by (\ref{506}) or by (\ref{507}). The
corresponding 4d solution (\ref{506}) is
\be  \label{4dextr}
\ba{l}
ds^2 = t_{-}^2  \left[ -d\eta^2 + d\Omega_{k}^2 \right] \\
\rho= \left( \frac{\sqrt{k}}{\tan \sqrt{k} \eta } \right) \qquad ,
\qquad e^{-2 \phi} \sim \cos \sqrt{k} \eta \, \frac{\sin \sqrt{k} \eta}
 {\sqrt{k}}  \ .
\ea
\ee
This extremal solution is valid for arbitrary $k$.  Especially, for
$k=0$ the dual solution describes the flat Minkowski space with vanishing
dilaton and linear modulus field.  The geometry for $k=1$ is $R \times
S^3$, i.e. a 4d throat and for $k=-1$ we have to replace $S^3$ by a 3d
pseudo sphere.

For $k=-1$ the other extremal limit coming from
(\ref{508}) yields as 4d theory
\be \label{4dextr2}
\ba{l}
ds^2 = \left\{1 + \frac{t_{-}^2}{\eta^2} \right\}
  \left[ -d\eta^2 + \eta^2 d\Omega_{k=-1}^2 \right] \\
\rho =1 \qquad , \qquad e^{2 \phi} = 1 + \frac{t_{-}^2}{\eta^2} \ .
\ea
\ee
The torsion in all cases is given by (\ref{506}). The geometry of
this limit is a half throat. For $\eta \rightarrow 0$ we reach again the
$R \times S_k^3$ geometry, whereas for $\eta \rightarrow \infty$ we end
up with a flat Minkowski space (see figure (2d)).  While in
(\ref{4dextr}) the (string) metric is static it is time dependent in
(\ref{4dextr2}). In the Einstein frame the situation is vice verse. After
performing the corresponding Weyl transformation we get for
(\ref{4dextr})
\be  \label{eextr}
ds_E^2 = \cos \sqrt{k} \eta \, \frac{\sin \sqrt{k} \eta}
 {\sqrt{k}} \left[ -d\eta^2 + d\Omega_{k}^2 \right]
\ee
which is again oscillating for $k=1$. For (\ref{4dextr2}) we obtain
a flat Minkowski space (the Weyl factor drops out) but a non trivial
dilaton.  Note that as long as $k \not= 0$ we have a non vanishing torsion.
%
%\begin{figure}[t] \vspace*{0mm} \par
%\hfill \begin{minipage}[t]{160mm}
%\mbox{\epsfig{file=bild2.eps,width=18cm}}
%   %bbllx=32pt,bblly=32pt,bburx=540pt,bbury=520pt}}
%      \vspace{-25mm} \par
%    \caption{In (a) we have plotted the closed oscillating solution
%for $k=1$ and (b) is the wormhole solution for $k=-1$. Below are the
%  both extremal limits: (c) corresponding to $R \times S_k^3$ and (d)
%the throat solution with one flat region for $k=-1$.}
%     \end{minipage}  \hfill
%\vspace*{15mm}
%\end{figure}
%-----------------------------------------------------------------
\begin{figure}[t] \vspace*{1mm}
%\begin{minipage}[t]{155mm}
\begin{tabular}{cc}
 \begin{minipage}[t]{8cm}
\mbox{\epsfig{file=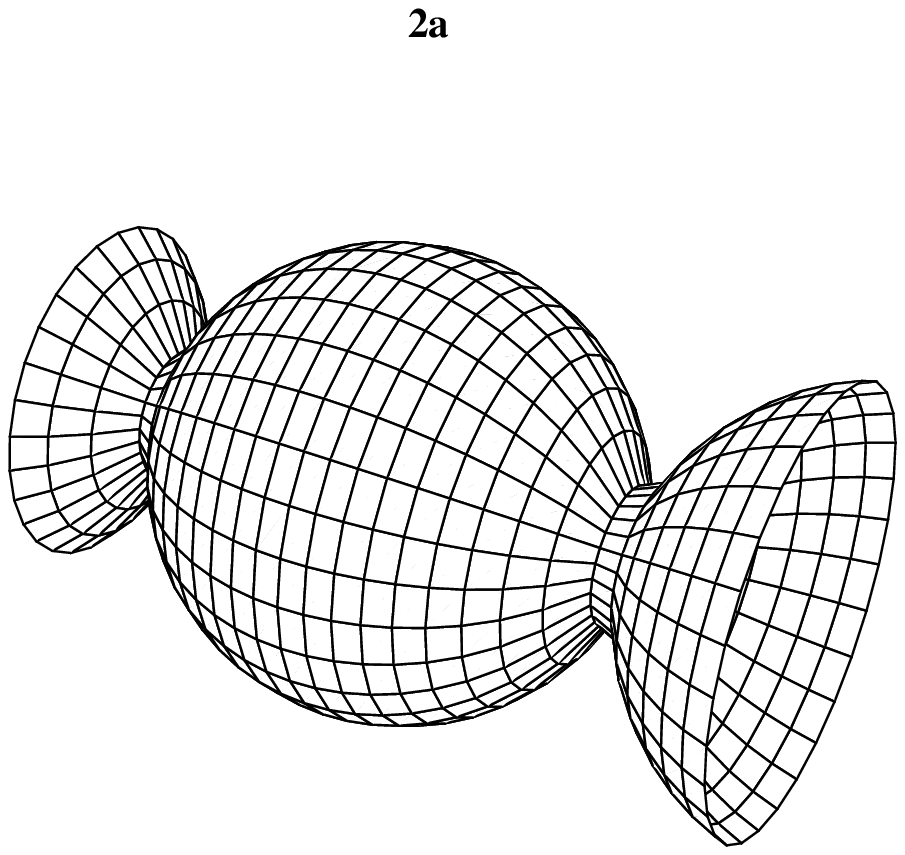,height=10cm,width=8cm}}
      \vspace{-27mm} \par
     \end{minipage}
 \hspace{-10mm} &
  \begin{minipage}[t]{8cm}
 \mbox{\epsfig{file=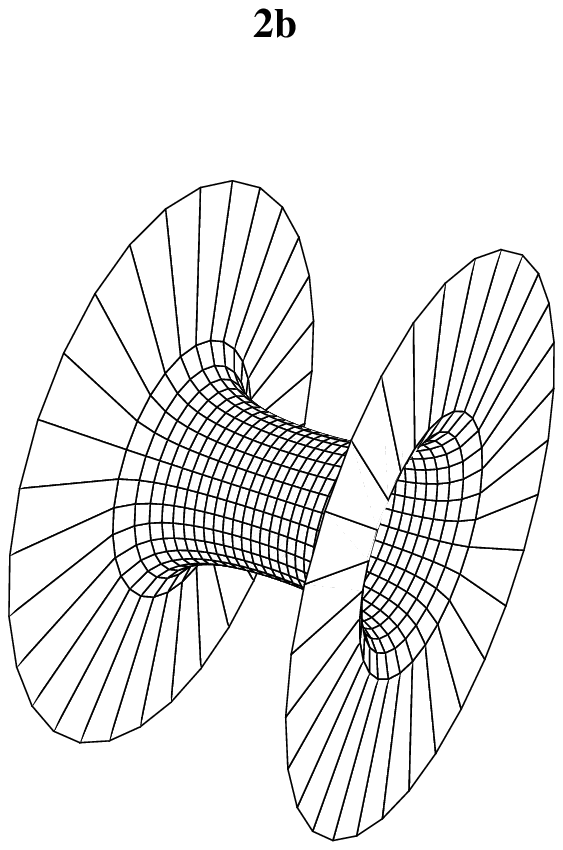,height=10cm,width=8cm}} \vspace{-27mm} \par
 \end{minipage}
\vspace{-1mm}\\
 \begin{minipage}[t]{8cm}
 \mbox{\epsfig{file=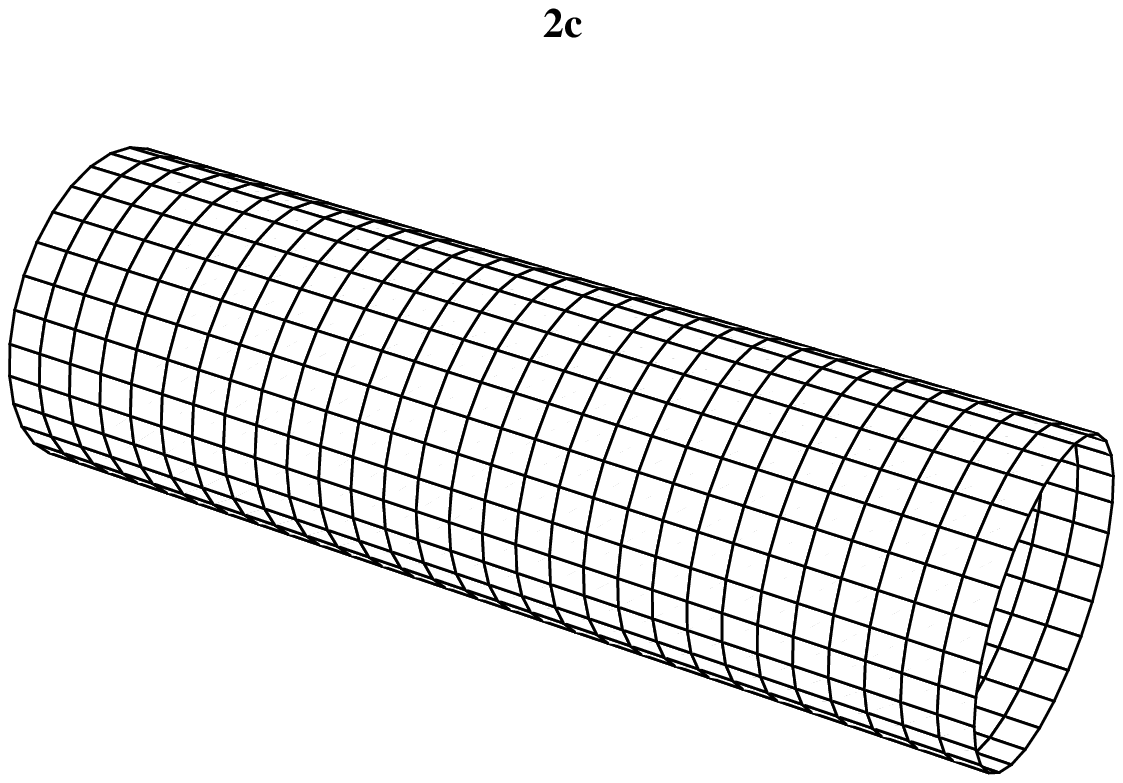,height=10cm,width=8cm}}
 \vspace{-27mm} \par
\end{minipage}
\hspace{0mm} &
 \begin{minipage}[t]{8cm}
 \mbox{\epsfig{file=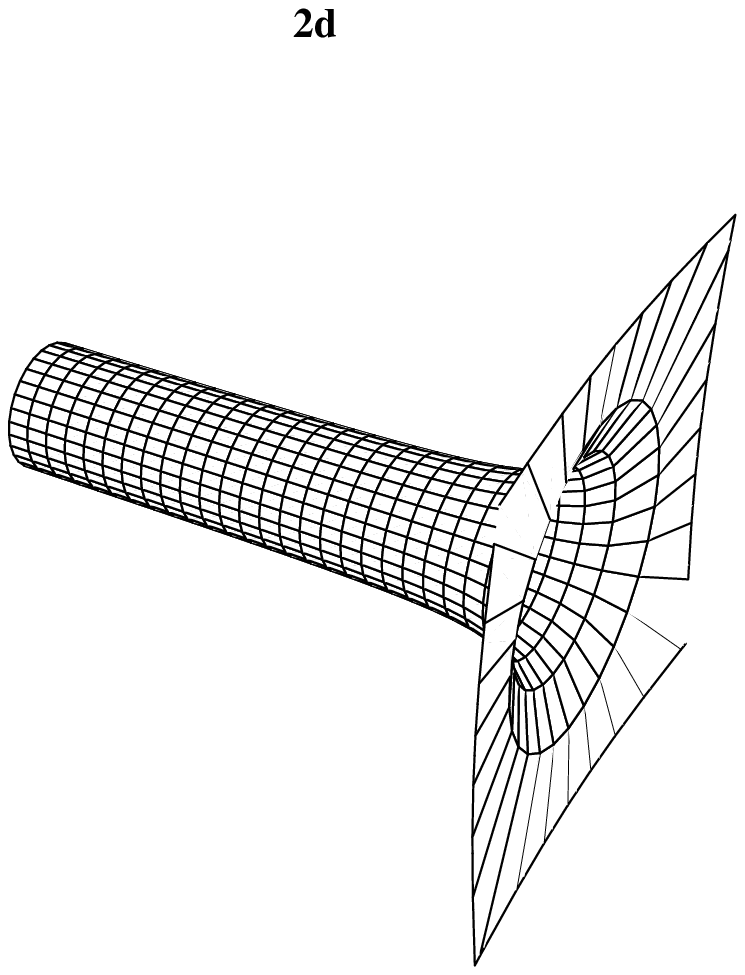,height=10cm,width=8cm}} \vspace{-27mm} \par
\label{bild2}
 \end{minipage}\vspace{-1mm}
\end{tabular}
\caption{In (a) we have plotted the closed oscillating solution
for $k=1$ and (b) is the wormhole solution for $k=-1$. Below are the
  both extremal limits: (c) corresponding to $R \times S_k^3$ and (d)
the throat solution with one flat region for $k=-1$. }
%\end{minipage}
\vspace{2mm}
\end{figure}
>From the geometry it is clear that both extremal limits are non
singular in the string frame. However, in the Einstein frame
(\ref{eextr}) is singular at certain points. If we calculate the Ricci
scalar in the extremal limit we find\footnote{ Note, that the extremal
limit in this case is given by (\ref{505}), i.e. the dilaton gets a
constant shift resulting in a constant Weyl transformation in the
metric.}
\be
R_E \sim \frac{1}{|\sin\sqrt{k} \eta \, \cos \sqrt{k} \eta|^3} \ .
\ee
Thus, corresponding to the zeros of the Weyl factor in (\ref{eextr})
the Ricci scalar has singularities. The asymptotic behavior of the
world radius in the proper time $\tau$ is: $a(\tau) \sim
\tau^{\frac{1}{3}}$ for $\tau\rightarrow0$ (independently of $k$) and at
infinity we obtain $a(\tau)\sim\sqrt{\tau}$ for $k=0$ and for $k=-1$
the metric (\ref{eextr}) is again asymptotically flat.

Finally, let us discuss what happens if we perform the extremal limit.
We have plotted the results in figures (2a-2d).  For $k=1$ in the
limit (\ref{505}) the maxima and minima approach each other yielding a
4d $R \times S^3$ geometry which is a static universe (figure (2c)).
In the case of $k=-1$ the solution has two asymptotic flat regions
connected by a wormhole (figure (2b)). In this case we have two
possibilities to perform the extremal limit. The limit (\ref{505})
shifts both flat regions to the infinite past or infinite future and
we have again a static universe with the only difference that $S^3$
has to be replaced by the 3d pseudo sphere. In the other extremal
limit one flat region remains fixed and the other one is shifted to
infinity. Looking on (\ref{4dextr2}) one could think that there are
two flat regions ($\eta \rightarrow \pm\infty$), but from every point
$\eta_0$ (positive or negative) the point $\eta=0$ is infinite
far away
\be
s = \int_{\eta_0}^{\eta} \sqrt{\tilde{\eta}^2 + t_-^2} \,
\frac{d\tilde{\eta}}{\tilde{\eta}} =
\left[ \sqrt{\tilde{\eta}^2 + t_-^2} - t_- \log
\frac{\sqrt{\tilde{\eta}^2 +t_-^2} + t_-}{|\tilde{\eta}|}
\right]_{\eta_0}^{\eta} \longrightarrow \infty \quad  \mbox{for} \quad
\eta \rightarrow 0
\ee
and therewith the other flat region is not reachable. Asymptotically
one can reach one flat region ($\eta\rightarrow\infty$ if $\eta_0 >0$;
see figure (2d)) or the point $\eta = 0$ (corresponding to minimal
extension of the throat). In both cases an infinite proper time is
necessary.  As long as we are not in the extremal limit the corresponding
length is finite.  Thus, the wormhole solution for $k=-1$ loses during
the extremal limit one or both of its flat regions.

\vspace{2mm}

\noindent {\bf Additional dust contribution:}
Finally, we want to discuss whether it is possible to include dust
matter in a consistent way. In \cite{wltsh,matzner} it was argued that
the 4d cosmological interpretation of a ($k=+1$) 5d back hole solution
is pathological. Namely, an additional dust contribution to the energy
momentum tensor would create a singularity near the maximal extension
of the universe. Let us investigate whether a similar effect occurs
here, too. For that reason we consider our 5d solution in the Einstein
frame and perform a time transformation to get the standard
Schwarzschild metric (\ref{3}) (note, that here the
functions $\lambda(t)$ and $\rho(t)$ do coincide with the solution
given in (\ref{5})).  An additional dust contribution to the 5d
energy momentum tensor is given by
\be  \label{dustpart}
T_{tt} = \mu(t) \, e^{\lambda} \ .
\ee
Assuming that for the dust part of the energy momentum tensor
the energy conservation is fulfilled
we find for the energy density
\be
\mu(t) =  \frac{C}{\rho \, t^3} \ .
\ee
Because the dust part of the energy momentum tensor contributes to
the $(t,t)$ part only the function $\lambda(t)$ remains unchanged
($\lambda$ is defined by the (5,5) component of the Einstein equation).
The $(t,t)$ part of the Einstein equation yields a modified
function $\rho(t)$
\be  \label{dust}
\rho = \left(1 + C\, \frac{2}{3} \int^t \frac{(1-\frac{1}{3}
  (\frac{t_-}{\tilde{t}})^2) d\tilde{t}}{[(-k +
  (\frac{t_+}{\tilde{t}})^2)(1 - (\frac{t_-}{\tilde{t}})^2)
  ]^{\frac{2}{3}} \, \tilde{t}^2} \right) \tilde{\rho}(t)
\ee
where $\tilde{\rho}$ is the compactification radius (modulus field)
when the dust contribution to the energy momentum tensor vanishes
($C=0$). We want to restrict ourselves on regions away from the point
of minimal extension ($\tilde{t}^2 = t_-^2$).  Obviously, the dust
contribution (\ref{dustpart}) is singular at zeros of $\rho$.  As long
as $C>0$ the zeros of $\rho$ coincide with the zeros of $\tilde{\rho}$
and the dust part yields no additional divergencies. But if $C<0$
additional zeros can occur. For $k=0,-1$ the integral is finite and as
long as $|C|$ is smaller than a critical value nothing disastrous
happens (it is reasonable to restrict oneself on small perturbations).
For $k=1$, however, the integral is divergent if we approach the
maximal extension of the 4d string metric $\tilde{t}^2 = t_+^2$ and
for all negative $C$ we get an additional zero of $\rho$.  Since this
point is a singularity or horizon of the 5d theory this behavior is
not surprising. Some confusion can appear after reduction to 4d
because the 4d string metric is completely smooth. But nevertheless at
this point the scalar matter part (dilaton and modulus) has
singularities and one can expect that also the dust matter part is
singular there.  In the 4d Einstein frame are no such shortcomings
because all zeros of $\rho$ and therewith singularities in the energy
momentum tensor are accompanied by curvature singularities. Sokolowski
\cite{soko} argued that just this behavior indicates that in this
context the Einstein frame is physically more reasonable.  Thus, we
can conclude that dust matter can be included consistently as long as
we are not too close to the extrema in the string frame or
equivalently not to close to the curvature singularities in the
Einstein frame.

\section{Conclusions}
In the present paper we discussed a combination of string and
Kaluza--Klein theory \cite{schwarz} in the context of cosmological
space time structures.  For that sake we had to generalize the 5d
black hole solution \cite{maeda,strom} to arbitrary constant curvature
of the spherical 3d subspace. Furthermore we had to
interchange the signature of time and radius in order to get a
cosmological solution after the Kaluza--Klein reduction.

We were able to show that the 5d solution generalized to arbitrary
constant spatial curvature possesses a limit where it is exact to all
orders in $\alpha^\prime$.  That was done in analogy to the
considerations in \cite{gidd}.  So, at least in a certain limit there
is an exact conformal field theory behind our 5d phenomenological
solution.

After performing the Kaluza--Klein reduction we got a four dimensional
configuration with an isotropic cosmological metric, a dilaton field,
torsion and a modulus field. Unfortunately there is no way to find an
analytical expression for the metric in the standard Robertson--Walker
form. That form is very suitable for the discussion of cosmological
scenarios. Therefore it is worthwhile to give a numerical solution for
the world radius. For certain special cases one can find analytical
results. However, in those special cases the dilaton and the torsion
of the 5d theory vanish and hence we get the result known from
Einstein gravity \cite{matzner}. Depending on the choice of
parameters we have finite or infinite universes. For $t_{\pm}^2>0$ our
cosmological solution is oscillating for $k=1$, for $k=-1$ we get a
wormhole solution with two flat regions (in the infinite past and in
the infinite future), for $k=0$ the geometry is also a wormhole but
without flat regions.  Transforming these solutions into the Einstein
frame has the consequence that the wormholes and all other extrema of
the string frame shrink to zero and form singularities. In some sense
the singularities in both frames are complementary: a singularity in
one frame is an (non singular) extremum in the other frame.  The
reason is, that the dilaton is divergent at all zeros and extrema of
the world radius.  In addition, we have briefly discussed the question
whether matter or information can pass the wormhole. In this region
the 5d metric has a smooth 3d spherical part and a singular 2d BH part
and for winding modes it is possible to pass the 2d BH singularity.
In addition to the numerical results the time dependence of the world
radius was given in some asymptotical regions.

Finally we discussed the question what happens when we throw dust into
the five dimensional space time. For that we transformed the 5d
effective string action to the Einstein frame and added a dust
contribution to Einstein's equations. We observed that for open
universes ($k=0,-1$) there will be no additional singularity in the
dust contribution as long as it does not exceed a critical value. For
closed universes ($k=1$) the dust contribution becomes singular near
the maximal extension of the universe in the string frame. But
because at this point the dilaton is divergent too, this singular
matter contribution is not surprising. Instead, transforming the
solution into the Einstein frame yields the result that all divergencies
in the matter contribution coincide with divergencies of the metric
(at the big bang or the big crunch).

The most interesting open question in our approach is whether it is
possible to find exact 5d (or higher dimensional) solutions which will
give a cosmological solution after Kaluza--Klein reduction. For closed
universes those exact solutions might be obtained by the consideration
of gauged WZW models consisting of a group which has the $SU(2)$ as a
subgroup. However, the gauging must not affect the $SU(2)$ subgroup in
order to preserve the $S_3$ geometry.  Perhaps those solutions will
also be interesting in the context of black hole physics, (as the one
we used obviously was).

\vspace{5mm}

\noindent
{\large\bf Acknowledgments}\vspace{3mm}\newline\noindent
S.\ F.\ would like to thank Amit Giveon and Gautam Sengupta for useful
discussions. K.\ B.\ is grateful to J.\ Garcia--Bellido, L.\ Dixon,
L.\ Susskind, A.A.\ Tseytlin and D.L.\ Wiltshire for helpful discussions
and remarks.

\renewcommand{\arraystretch}{1.0}

\end{document}